# Single crystal synthesis and low-lying electronic structure of V$_3$S$_4$

Yu-Jie Hao[1,2,*], Ming-Yuan Zhu[2,*], Xiao-Ming Ma[2*], Chengcheng Zhang[2], Hongtao Rong[2], Qi Jiang[3], Yichen Yang[3], Zhicheng Jiang[3], Xiang-Rui Liu[2], Yupeng Zhu[2], Meng Zeng[2], Ruie Lu[4], Tianhao Shao[2], Xin Liu[2], Hu Xu[2], Zhengtai Liu[3], Mao Ye[3], Dawei Shen[3], Chaoyu Chen[2], Chang Liu[2†]

[1]*Department of Physics, Harbin Institute of Technology, Harbin, Heilongjiang 150001, China*

[2]*Shenzhen Institute for Quantum Science and Engineering (SIQSE) and Department of Physics, Southern University of Science and Technology (SUSTech), Shenzhen, Guangdong 518055, China*

[3]*State Key Laboratory of Functional Materials for Informatics, Shanghai Institute of Microsystem and Information Technology, Chinese Academy of Sciences, Shanghai 200050, China*

[4]*School of Mechanical and Electrical Engineering, Guangzhou University, Guangzhou 510006, China*

*These authors contributed equally to this work.
†Corresponding author. E-mail: liuc@sustech.edu.cn

**Abstract**

We report successful growth of millimeter-sized high quality single crystals of V$_3$S$_4$, a candidate topological semimetal belonging to a low-symmetry space group and consisting of only low atomic number elements. Using density functional theory calculations and angle-resolved photoemission spectroscopy, we show that the nonmagnetic phase of monoclinic V$_3$S$_4$ hosts type-II Dirac-like quasiparticles which opens a sizable gap due to spin orbit coupling, as well as theoretical multiple nodal lines that are eliminated also by spin orbit coupling. These results suggest that relativistic effects give rise to profound modifications of the topological properties even in compounds with low-weight elements.

## 1. Introduction

Topological semimetals (TSMs) received a frenzy of research activity due to the promise of exotic physical phenomena such as Fermi arcs and chiral anomalies [1], and potential applications in low-dissipation transport and quantum computation [2]. Differentiated by the governing equation of low-lying electronic states and the associated quasiparticles, topological semimetals can be classified as (but not limit to) Dirac semimetals [3, 4], Weyl semimetals [5-7], and nodal-line semimetals [8-11]. According to the sign of the tilt angles of the cones, Dirac / Weyl / nodal-line semimetals can further be classified as type-I or type-II [12-16]. The influence of spin orbit coupling (SOC) on the electronic properties of topological materials is a widely concerned topic [16-18]. In the presence of SOC, according to crystal symmetry and the SOC strength which grows rapidly with the atomic numbers of the constituent elements, some nodal states hold an experimentally negligible band gap (e.g., ZrSiS) [9, 19], while other band crossings will be gapped, changing the systems to trivial insulators or materials of other topological classes [20-23]. In the past years, many different types of TSMs have been experimentally verified [16-18]. Among these systems, compounds with a monoclinic structure receive less scientific attention, partly due to the difficulties in single crystals synthesis and the complicated band structures associated with the tilted Brillouin zone (BZ).

$V_3S_4$ is one of the representative compounds in three-dimensional topological-related materials with a low space group symmetry. It has a monoclinic $Cr_3S_4$-type crystal structure with space group C2/*m*. The Néel temperature ($T_N$ ~ 9 K) is verified by nuclear magnetic resonance [24], though results from vibrating sample magnetometer (VSM)-based quantum transport show no evidence of long range magnetic order. $V_3S_4$ is predicted as a topological semimetal via the method of topological quantum chemistry and symmetry indicators [25-28], as well as a Weyl orbital semimetal where symmetry-protected Weyl cones exist in the bulk bands without the need of SOC [29]. $V_3S_4$ also attract much attention by exhibiting a high theoretical capacity and low reaction potential as an anode material in potassium ion batteries [30]. However, its experimental electronic structure has not been systematically investigated to the best of our knowledge.

In this work, we report a method of synthesizing high quality, large size monoclinic $V_3S_4$ single crystals by combining direct reaction and chemical vapor transport (CVT) techniques [31]. Using first-principles density functional theory (DFT), we calculate the band structure of $V_3S_4$ in the nonmagnetic phase, without and with SOC. When neglecting SOC, a pair of tilted bulk Dirac cones locate along the Γ-Y direction in the Brillouin zone, and several topological nodal lines appear in the BZ near the Fermi level. When including SOC, the tilted Dirac cone opens a ~120 meV gap, and the nodal lines disappear. Our results from angle-resolved photoemission spectroscopy (ARPES) agree well with the DFT bands including SOC. Therefore, the existence of SOC affect greatly the bands of $V_3S_4$, a compound consists of only low-weight elements.

## 2. Experimental method

*Material Synthesis*: Single crystals of $V_3S_4$ were grown by the chemical vapor transport (CVT) method using high-purity $V_3S_4$ powder as the starting material with $I_2$ as the transport agent.

Synthesis of powder. – Polycrystalline $V_3S_4$ is produced by a solid state reaction method. V powder (99.9%) and S powder (99.99%) were finely grounded and mixed with a ratio of 44:56 [32, 33]. The mixture was then pressed into a pellet of approximately 12 MPa, which was put into a $SiO_2$ ampoule, evacuated to at least 10 mTorr and backfilled with 1/5 partial atmosphere of Ar. The sealed ampoule was placed in a tube furnace, heated to 450 °C and held for 5 hours, then heated to 1150 °C and held for an additional 5 hours (heating ramp 14 °C/h). Finally, it was cooled down to room temperature.

Single crystal growth by chemical vapor transport. – Single crystals of $V_3S_4$ were grown by the chemical vapor transport (CVT) with $I_2$ as a transporting agent. 1 g of the abovementioned finely grounded polycrystalline $V_3S_4$, checked with XRD, was sealed in an evacuated quartz ampoule (22 mm in diameter and about 250 mm long) after the addition of 8 mg/ml of $I_2$. The sealed ampoule was heated in a two-zone furnace to a low temperature $T_L$ = 800 °C and a high temperature $T_H$ = 900 °C in 60 hours. After maintaining at this condition for two months, the sealed ampoule was cooled down to room temperature. Using this procedure, shiny single crystals of sizes up to 3 × 2 × 0.5 $mm^3$ were obtained, as shown in the inset of Fig. 1(e).

*PPMS Measurements:* Resistivity measurements were performed at temperatures between 2 K and 300 K, and magnetic fields between -14 T to 14 T, in a Quantum Design Physical Properties Measurement System (PPMS) by a standard four-probe technique, with a drive current of 4 mA.

*ARPES Measurements:* ARPES measurements were performed at Beamline 03U of the Shanghai Synchrotron Radiation Facility (SSRF) [34, 35], Shanghai, China. The energy and angular resolutions were set to 10 meV and 0.1°, respectively. The samples were cleaved *in situ* along the (100) plane and measured at $T \sim 15$ K (nonmagnetic phase of $V_3S_4$) in a working vacuum better than $5\times10^{-11}$ Torr. Additional ARPES data (not shown) was also collected at Beamline 09U (Dreamline) of SSRF. The ARPES images obtained in rapid scanning modes are corrected by a Fourier-based method to eliminate the artifact caused by a mesh in front of the detector [36].

*First-Principles Calculation:* The electronic structure calculations were done using the DFT method encoded in the Vienna Ab-initio Simulation Package (VASP) [37, 38]. The core electrons were described by the projector-augmented wave (PAW) pseudo-potentials [39]. The Perdew-Burke-Ernzerhof (PBE) approximation is used for the exchange-correlation function [40]. The DFT-D3 method was also used to include the van-der-Waals correction [41, 42]. Plane waves with a kinetic energy cutoff of 520 eV were used as the basis set. The *k*-point sampling is $11 \times 11 \times 6$ with the Γ scheme. Both cell parameters and atomic positions are fully relaxed until the forces on each atom are smaller than $10^{-3}$ eV Å$^{-1}$, and the total energy convergence criterion is set to be $10^{-7}$ eV Å$^{-1}$. GGA+*U* correction is applied to the V 3*d* orbitals, and *U* is set to be 2.5 eV. The five V *d* orbitals and three S *p* orbitals are used to construct the maximally localized Wannier functions [43]. The nodal lines were calculated using the WANNIERTOOLS package, where the gap threshold was set as $10^{-4}$ eV when searching for the gapless points [44]. The representation analysis was carried out with the Irvsp code together with the VASP package [45].

## 3. Results and discussion

High quality, millimeter-sized $V_3S_4$ single crystals are grown via the chemical vapor transport (CVT) method with $I_2$ as the transport agent, shown in Fig. 1(a). The crystal structure and schematic bulk BZ of $V_3S_4$ is shown in Figs. 1(b)-(c). The core-level photoemission spectrum of $V_3S_4$ is shown in Fig. 1(d), from which the characteristic peaks of V and S elements are clearly observed. The peak at 120 eV belongs to the iodine transport agent. Fig. 1(e) shows the single crystal / powder x-ray diffraction data of $V_3S_4$. The peaks show a perfect match with the reported peak positions for the monoclinic C2/*m* structure of $V_3S_4$ [ICSD Code 79969], and no impurity phases are observed. The single crystals, sized up to 3 × 2 × 0.5 mm$^3$, show a typical morphology with clear surfaces, which are confirmed to be the natural (*l*00) planes. Lattice parameters of *a* = 12.616 Å, *b* = 3.286 Å, *c* = 5.870 Å and unit cell angles of *α* = 90°, *β* = 115.691°, *γ* = 90° are obtained by refining the powder x-ray diffraction data. Transport and magnetic properties of $V_3S_4$ are further depicted in Fig. S1. The ground state magnetic structure of $V_3S_4$ is currently under debate. Previous magnetic susceptibility measurements on $V_3S_4$ show no evidence of long range magnetic order [24, 46, 47], while a Néel temperature of about 9 K is found by a $^{51}$V nuclear magnetic resonance experiment [24, 48]. A ferromagnetic (FM) ground state below 4.2 K is also suggested, due possibly to an impurity phase [24]. Result of our magnetic susceptibility measurements on a large-size $V_3S_4$ single crystal is shown in Fig. S1(c). We do not see obvious antiferromagnetic characteristics, and the magnetization value (*M* ~ 10$^{-3}$ $\mu_B$/f.u.) is extremely small.

Next we discuss the electronic band structure of $V_3S_4$, as revealed by our ARPES measurements and DFT calculations. First, we demonstrate in Figure 2 that nonmagnetic $V_3S_4$ hosts a pair of bulk, tilted Dirac cones along the Y-Γ-Y direction, but they are not robust against the inclusion of SOC. Before presenting the band structure, we note that the BZ of a monoclinic crystal [Fig. 2(a)] is misaligned in the reciprocal space. In the case of $V_3S_4$, it is not until the Γ point of the 12$^{th}$ BZ along the $k_y$ direction that the same $k_z$ value can approximately be found with the Γ point of the first BZ [Fig. S2(a)]. Detailed procedure for judging the in-plane high symmetry points is shown in Figs. S2-S3. Keeping these complexities in mind, we found that, when SOC is turned off, the dispersion around the Dirac point is tilted along Γ-Y but not the two other orthogonal *k* directions, signaling the existence of Lorentz-violating type-II Dirac fermions. The bulk BZ and the projected (001) surface BZ are shown in Fig. 2(a) where high-symmetry points, lines and bulk Dirac points (labeled as D) are indicated. Fig. 2(b) shows the band structure of $V_3S_4$ measured at $h\nu$ = 245 eV, close to one of the D points. The 3D plot shows a "V" like band and a "Λ" like band along both $k_x$ and $k_y$ directions at D, confirming its in-plane Dirac-like dispersion with a gap. A DFT-derived three-dimensional band structure without SOC on the $k_x$-$k_z$ plane is shown in Fig. S4 where the tilted Dirac cone can clearly be seen. From the representation analysis, this band crossing is found unavoidable, because the two bands belong to different representations ($\Delta_1$ and $\Delta_2$), respectively, as distinguished by the $\sigma_h$ mirror symmetry in the Z-Γ-Y plane. Although the $\Delta_1$ and $\Delta_2$ representations are one-dimensional, such that the degeneracy is an accidental one, the difference of the representations does allow the bands crossing at a single nodal point. However, when SOC is included, the SU(2) symmetry is broken, an energy gap will open at the type-II Dirac points along Γ-Y. These results match well with our measured spectra, shown in Fig. 2(b). In order to ascertain the gap inducing by SOC, we calculated in Fig. 2(c) the in-plane band structures at different $k_z$s [blue lines in Fig. 2(a)] without / with SOC, where $k_z = 0$ and $k_z = 1$ define the bulk Γ and Y points, respectively. At $k_z = 0.49$, SOC-free calculations show a gapless Dirac cone at the zone center marking with "DP". When SOC is included, a ~120 meV gap opens at the cone. Detailed electronic structure along the high symmetry directions is presented in Fig. S5. As we all know, SOC strength scales as $Z^4$ ($Z$ being the atomic number). Although the atomic numbers of V and S are relatively small, the SOC in this system has a great influence on the electronic band structure. [49, 50] ARPES $E$-$k$ maps measured at several representative photon energies are shown in Fig. 2(d) (The relationship between the photon energy and the $k$-space position is confirmed by analyzing the ARPES $k_z$ dispersion map shown in Fig. S6). One clearly observes the Dirac-like band with a ~120 meV gap at $k_z = 0.52$ (close to D), and the gap situates at about 1.4 eV below the Fermi level. The gap size further increases when the photon energy decreases, and the largest gap size is about 450 meV at $h\nu = 230$ eV ($k_z = 0.95$). Fig. 2(e) shows the DFT band structure along Γ-Y with SOC. Green circles / yellow triangles represent the bottom / top of the gap in Fig. 2(d). A reasonable match is found between the ARPES and DFT results, which further confirms a gapped, type-II Dirac like band structure.

Second, we demonstrate in Figure 3 that nonmagnetic $V_3S_4$ is originally a topological nodal-line semimetal, whose nodal lines are gapped by SOC. When SOC is off, multiple topological nodal lines (NLS) exist close to $E_F$; when SOC is on, all of these nodal points are gapped. We calculated the nodal lines in the 3D BZ. The $k$-space span of the NLS1-NLS3 nodal lines viewed along $k_z$ ($k_x$) are shown in Figs. 3(a), 3(c) and 3(e) [Figs. 3(b), 3(d) and 3(f)]. NLS1 / NLS2 / NLS3 is formed by the crossing point of the 14$^{th}$ and 15$^{th}$ / 15$^{th}$ and 16$^{th}$ / 16$^{th}$ and 17$^{th}$ band [Fig. 3(g)]. The DFT band structure along the I-Z-I direction without SOC is shown in Fig. 3(g), in which the three band-touching points belonging respectively to NLS1-NLS3 are labeled as N$_1$, N$_2$, and N$_3$. According to the orbital character analysis, the N$_1$, N$_2$ and N$_3$ points are provided by the $d$ orbitals of the two different V atoms in the unit cell, preserving the time-reversal symmetry $T$. The point group of the V$_3$S$_4$ structure is $C_{2h}$, preserving the space inversion $P$. Therefore, the three nodal lines are protected by the combined symmetry $PT$ in the absence of SOC. The gapless nature of NLS1-NLS3 is found by first identifying all the $k$ points with zero energy gap within the first BZ using the WANNIERTOOLS package [44]. The topological protection of these nodal lines can then be inferred from calculating the topological number of the form $\gamma = \oint_C \mathbf{A}(\mathbf{k}) \cdot d\mathbf{k}$, where $\mathbf{A}(\mathbf{k}) = -i\sum_{n\in occ.}\langle u_n(\mathbf{k})|\partial_{\mathbf{k}}|u_n(\mathbf{k})\rangle$ is the Berry connection of the occupied states, $C$ is a closed loop in the momentum space. If $C$ is pierced by a nodal line, one has $\gamma = \pi$, otherwise $\gamma = 0$. We calculated the Berry phase with a loop around each nodal line, and always obtain $\gamma = \pi$. Therefore, the nodal lines presented in Figs. 3(a)-(f) are topologically nontrivial. When SOC is included, the SU(2) symmetry will be broken, so all of these crossing bands will be gapped and V$_3$S$_4$ will become topologically trivial. For the band structure along the I-Z-I direction [Fig. 3(h)], the gap size in one of the original nodal points (marked as N') is calculated to be ~10 meV, as shown in the inset of Fig. 3(h). This small gap is invisible in our ARPES data due to limited energy resolution. In Fig. 3(i) we show the ARPES band dispersion along I-Z-I appended by DFT band dispersions with SOC (red lines). The ARPES spectrum matches well with the SOC-included bands.

For completeness of our discussion, we also checked the ground energies of V$_3$S$_4$ by VASP. The ground energy of V$_3$S$_4$ with FM order is -44.5496 eV. The ground energy of V$_3$S$_4$ with NM order is -42.3278 eV, which is higher than the FM order. Therefore, an *ab initio* study on the band structure in the FM state is necessary. In the Supporting Information, we performed DFT band calculations of V$_3$S$_4$ in the FM state, in which each V atom has a magnetic moment of about 2.1 $\mu_B$. Our results showed that it is a magnetic Weyl semimetal [51]. In the FM state, the spin-down part is an insulator with a large direct band gap near the Fermi surface in the whole

BZ, while the spin-up part forms a Weyl point along the I-Z direction (Fig. S7). The orbital components near the Weyl point are basically provided by the *d* orbitals of the two unequal V atoms. The Weyl point appears at the Fermi level due to the number of electrons occupied, as we can see in the electronic structure (Fig. S8). It is noted, however, that although the ground state energy of the FM state is calculated to be lower than that of the nonmagnetic state, this FM ground state is not observed in our transport and magnetic measurements [Fig. S1(c)].

## 4. Conclusion

In summary, we demonstrate the growth of millimeter-sized high-quality single crystals of $V_3S_4$ using the chemical vapor transfer method. We systematically study the electronic band structure of $V_3S_4$ via ARPES measurements and DFT calculations. When SOC is not included, we find that the nonmagnetic phase of $V_3S_4$ hosts both a type-II Dirac semimetallic state and a nodal-line semimetallic state, with a pair of tilted Dirac cones at the Γ-Y direction and several prolonged nodal lines near the Fermi level, which are protected by a combination of *P* and *T* symmetries. However, when SOC is included, these bands are gapped. The Dirac-cone-like dispersion opens a ~120 meV gap, and the nodal lines vanish everywhere in the BZ. Therefore, the energy bands of $V_3S_4$ changed greatly under the action of SOC. Since the effect of SOC is nearly negligible in compounds like $SrAs_3$ and $EuAs_3$ which shares the same C2/*m* space group as $V_3S_4$ and consisting of even heavier elements, it is likely that the underlying mechanism that determines how much SOC would affect the electronic structure of a topological material might be different in these systems. As a concluding remark, it is pointed out in the literature that SOC tends to induce rather than eliminate type-II Dirac points in selected systems. In $PtSe_2$ and $VAl_3$, when SOC is turned on, a triple degenerate point becomes a type-II Dirac point [14, 15]. In AMgBi ( A = K, Rb, Cs), SOC effect causes band inversion to form a type-II Dirac point [52]. In $NiTe_2$ and $CaAl_2Si_2$, due to the crystal field splitting and SOC effect, the *p* orbital splits and forms a type-II Dirac point [53, 54]. In light of these results, and together with our observations, one realizes that the influence of SOC is two-folded and complicated. In the past, strong SOC effect was mainly considered in materials with heavy elements. In this work, we have observed that there are also pronounced SOC effects in light element materials, indicating that even light element materials may contain a wealth of exotic quantum phases such as a topological insulator with non-toxic, harmless and stable constituents.

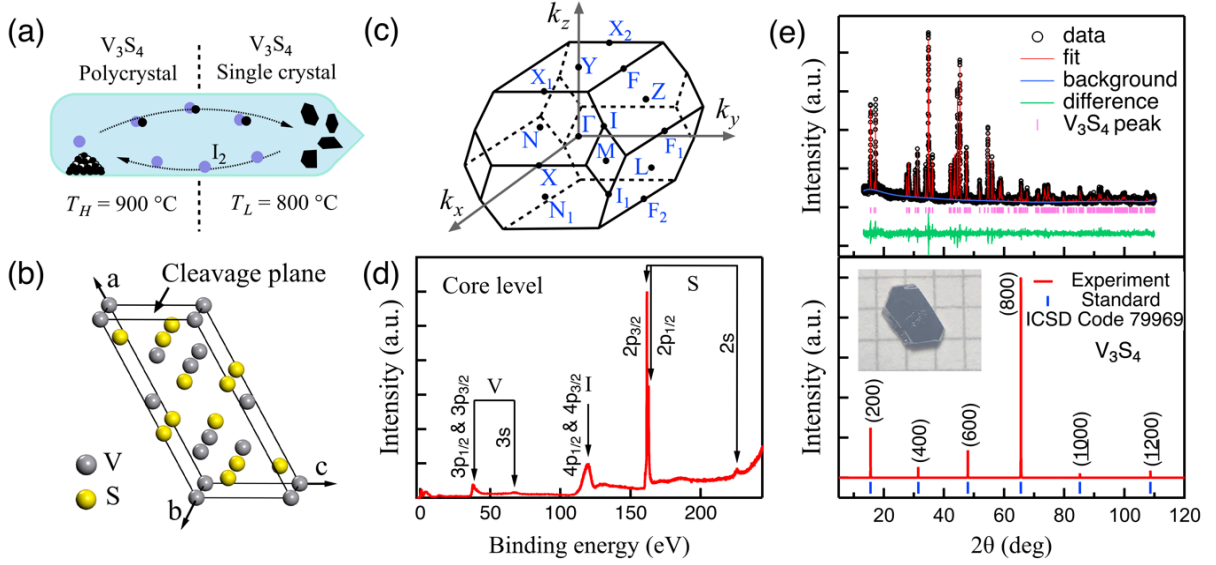

**FIG. 1** (a) Schematics of the CVT growth of $V_3S_4$ single crystals. (b) Crystal structure of $V_3S_4$. (c) 3D Brillouin zone of $V_3S_4$, with definitions of the $k$ axes. (d) Core level photoemission spectrum, showing the $p$-shell peaks of V, S, and the transport agent $I_2$. (e) Top: powder x-ray diffraction (XRD) pattern of $V_3S_4$. Open circles: XRD data; red line: fit to the data; blue line: background of data; green line: difference between the data and the fit; pink vertical bars: XRD peak positions of $V_3S_4$. The refined lattice parameters are $a$ = 12.6159(2) Å, $b$ = 3.2862(1) Å, $c$ = 5.8705(1) Å; unit cell angles are $\alpha = 90°$, $\beta = 115.691(1)°$, $\gamma = 90°$. Bottom: single crystal XRD pattern taken at 300 K. Inset shows a typical single crystal against a millimeter grid.

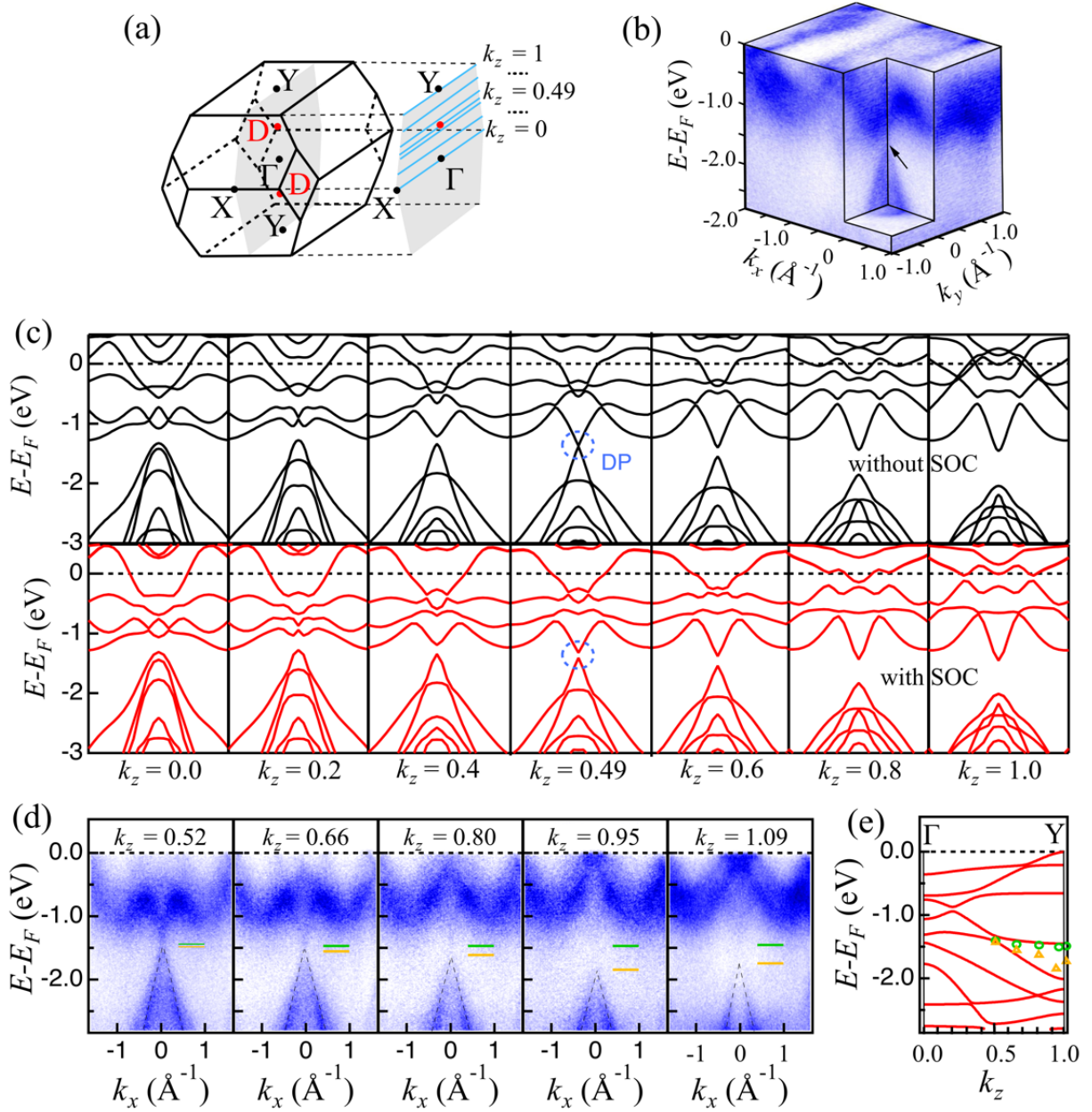

**FIG. 2** (a) Bulk and projected surface Brillouin zone (BZ) and definition of the high-symmetry points. Red dots (labeled as D) mark the positions of the gapped Dirac points. $k_z$ values are defined in units of $k_{\Gamma\text{-Y}}$. (b) 3D intensity plot of ARPES spectra measured at $h\nu = 245$ eV ($k_z = 0.52$, close to the bulk D point). The gapped Dirac cone is marked by an arrow. (c) DFT band dispersions along an in-plane direction parallel to X-Γ-X without/with SOC, calculated at different $k_z$ values in the reduced BZ [blue lines in (a)]. When SOC is included, the Dirac cone opens a ~120 meV gap. (d) ARPES band dispersions along X-Γ-X, taken at $h\nu = 245$ eV ($k_z = 0.52$), 240 eV ($k_z = 0.66$), 235 eV ($k_z = 0.80$), 230 eV ($k_z = 0.95$), and 225 eV ($k_z = 1.09$). Evolution of the band dispersion from the D point to the Y point is seen. Green/orange lines indicate the top/bottom of the gap. (e) DFT band structure along Γ-Y with SOC. Green circles / orange triangles represent the experimentally determined top/bottom of the gap in (d). The gap is seen to enlarge from D to Y.

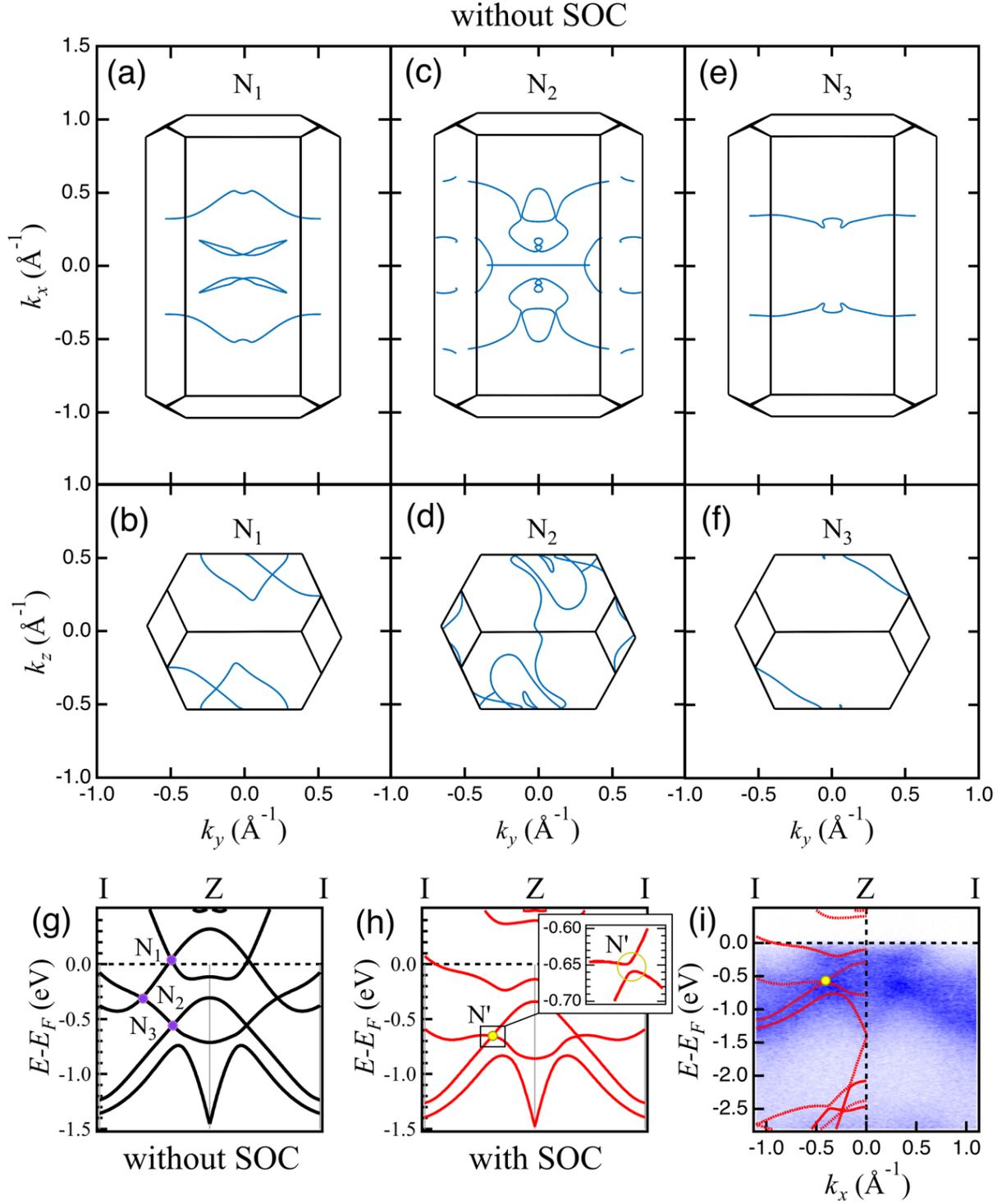

**FIG. 3** (a)-(f) *k*-space span of the nodal lines viewed along $k_z$ [(a), (c), (e)] and $k_x$ [(b), (d), (f)]. (g) DFT band structure along I-Z-I without SOC. The three nodal line intersections are marked as $N_1$, $N_2$ and $N_3$. (h) DFT band structure along I-Z-I with SOC. Inset: enlarged view near the N' point, emphasizing the presence of an energy gap. (i) ARPES spectra taken along I-Z-I with 245 eV photons, appended with DFT band dispersions (red lines).

**CRediT authorship contribution statement**

C.L. conceived and designed the research project. Y.-J.H. and X.-M.M. grew and characterized the single crystals. Y.-J.H., X.-M.M. and R.L. performed the transport and magnetic measurements with help from T.S. Y.-J.H., X.-M.M., C.Z., H.R., X.-R.L., Y.Z., M.Z., R.L., T.S., X.L., C.C., and C.L. performed the ARPES measurements with onsite support from Q.J., Y.Y., Z.J., Z.L., M.Y., and D.S. M.-Y.Z. performed the DFT calculations with help from H.X. Y.-J.H., M.-Y.Z., and C.L. analyzed the data and wrote the paper.

**Data Availability**

Data are available from the corresponding author upon reasonable request.

**Declaration of Competing Interest**

The authors declare no competing financial or non-financial interests.

**Acknowledgments**

Y.-J. H., M.-Y. Z. and X.-M. M. contributed equally to this work. We thank Peipei Wang and Liyuan Zhang for help in the CVT growth of the single crystals. We thank Yuanjun Jin, Pengfei Liu and Qihang Liu for helpful discussions. Work at SUSTech was supported by the National Natural Science Foundation of China (NSFC) (Nos. 12074161, 12074163, 11804144, 11504159), NSFC Guangdong (Nos. 2022A1515012283, 2016A030313650), and the Guangdong Innovative and Entrepreneurial Research Team Program (No. 2016ZT06D348). The ARPES experiments were performed at BL03U of Shanghai Synchrotron Radiation Facility under the approval of the Proposal Assessing Committee of SiP.ME$^2$ platform project (Proposal No. 11227902) supported by NSFC. D. S. acknowledges support from NSFC (No. U2032208). C. C. is supported by Guangdong Basic and Applied Basic Research Foundation (Grants No. 2022B1515020046 and No. 2021B1515130007). C. L. acknowledges support from the Highlight Project (No. PHYS-HL-2020-1) of the College of Science, SUSTech.

**Appendix A. Supporting information**

Supplementary data associated with this article can be found in the online version.